%% file: DC-aps.tex
\documentclass[
preprint,
 amsmath,amssymb,
 aps,
]{revtex4-1}

\usepackage{graphicx}
\usepackage{dcolumn}
\usepackage{bm}

\def\d{\mbox{\rm d}}

\def\e{\mbox{\rm e}}
\def\i{\mbox{\rm i}}
\def\be{\begin{eqnarray}}
\def\ee{\end{eqnarray}}
\def\bme{\mbox{\boldmath $e$}}
\def\bmU{\mbox{\boldmath $U$}}
\def\bmV{\mbox{\boldmath $V$}}
\def\bmv{\mbox{\boldmath $v$}}
\def\smalli{\mbox{\scriptsize \rm i}}
  \makeatletter
    \renewcommand{\theequation}{%
    \thesection.\arabic{equation}}
    \@addtoreset{equation}{section}
  \makeatother

\begin{document}


\title{Drag coefficient of a liquid domain in a fluid membrane \\with the membrane viscosities being different\\ across the domain perimeter}

\author{Hisasi Tani}
 \email{hisasitani@gmail.com}
 \affiliation{Organization for the Strategic Coordination of Research and Intellectual Properties, Meiji University,
Kawasaki 214-8571, Kanagawa, Japan}
\author{Youhei Fujitani}%
\affiliation{%
 School of Fundamental Science and Technology, Keio University, Yokohama 223-8522, Kanagawa, Japan
}%




\date{\today}

\begin{abstract}
We calculate the drag coefficient of a liquid domain in a flat fluid membrane surrounded by three-dimensional fluids
on both sides.  
In the membrane, the tangential stress should be continuous across the domain perimeter, 
which makes the velocity gradient discontinuous there unless the ratio of the membrane viscosity inside the domain to 
the one outside the domain equals unity.  
The gradient of the velocity field 
in the three-dimensional fluids is continuous.   
This field, in the limit that the spatial point approaches the membrane, should agree with 
the velocity field of the membrane.   Thus, unless the ratio of the membrane viscosities is unity,
we need to assume some additional singularity at the domain perimeter in solving the governing equations. 
In our result, the drag coefficient
is given in a series expansion with respect to a dimensionless parameter, 
which equals zero when the ratio of the membrane viscosities is unity and approaches unity when the ratio tends to infinity.
We derive the recurrence equations for the coefficients of the series.
In the limit of the infinite ratio, our numerical results agree with the previous results for the disk.   
\end{abstract}

\maketitle


\section{Introduction \label{sec:intro}}
The magnitude of the drag force exerted on a colloidal particle moving slowly enough in a fluid is proportional to its speed. 
The constant of the proportion is called drag coefficient, 
of which reciprocal gives the diffusion coefficient after being multiplied by the Boltzmann constant and temperature 
\citep{Sutherland, Einstein}.  
Calculating the drag coefficient is one of the fundamental problems in the low-Reynolds number hydrodynamics \citep{HB}.   
Most well-known is the drag coefficient of a rigid sphere in a three dimensional (3D)
fluid \citep{Stokes}.  
That of a droplet is also well known \citep{Hadamard, Rybczynski}.  
The latter tends to the former as the ratio of the viscosity of the droplet to that of the ambient fluid approaches infinity.
It is to be noted that we need not consider droplet deformation in this linear regime.  \\

One can neglect the inertia term to use {the} Stokes equation when the Reynolds number is small.
The drag coefficient of a disk in a two-dimensional (2D) fluid cannot be calculated in the Stokes approximation \citep{Lamb}. 
This Stokes paradox can be helped when the fluid is sandwiched by 3D fluids.  
This situation, for example, is realized by using a lipid-bilayer membrane, which is the main part 
of biomembrane and has the fluidity \citep{Singer}.
The drag coefficient was calculated for a small disk in a flat fluid membrane surrounded by 3D fluids occupying semi-infinite spaces 
on both sides \citep{SD, Saffman}; the result was utilized experimentally (\citet{PetersCherry}).
Theoretically, in this geometry, we use the cylindrical coordinates to introduce the
Hankel transformation. 
In the end, we need to solve a set of integral equations, 
which were studied extensively \citep{Sneddon, Hughes}.  \\

The main lipid component of the biomembrane is phospholipid.  
Some minor lipid components are concentrated to form a liquid 
domain called a lipid raft, which is ten to several hundreds nanometer in size \citep{PartonSimons, SimonsToomre, SubczynskiKusumi}.
It is thought to play significant roles in biological activities, 
for instance, in the signal transduction.
Raft-like liquid domains in an artificial fluid membrane have also 
been studied in the context of phase separation \citep{Veatch, Yanagisawa}. 
The drag coefficient of a 
liquid domain whose membrane viscosity $(\eta_{\rm i})$ equals
 the one outside the domain $(\eta_{\rm o})$ was calculated 
in \citet{DeKoker}.   Here, we introduce a dimensionless parameter defined as
\begin{equation}
	\kappa \equiv 1 -{\eta_{\rm o} \over \eta_{\rm i}} \ .\label{kappa}
\end{equation}
In  \citet{DeKoker}, $\kappa$ is supposed to vanish.
Cases where the membrane viscosities are slightly different were studied  in \citet{Fujitani80}, where
the drag coefficient is calculated up to a linear order of $\kappa$.
However, as described later, one boundary condition is overlooked there.  This error is corrected 
and the drag coefficient is recalculated up to the same order in \citet{Fujitani82}.
In the present study, using the corrected boundary condition, 
we calculate the drag coefficient of a liquid domain
by considering terms of higher order with respect to $\kappa$.
In particular,  as $\kappa$ approaches unity, {\it i.e.\/}, as $\eta_{\rm o}$ becomes much larger than $\eta_{\rm i}$, 
our result successfully
tends to the drag coefficient of a disk, which is calculated by the formula obtained in \citet{Hughes}. 
A related work is found in \citet{RD}, where the uncorrected boundary condition of \citet{Fujitani80} were 
somehow used although \citet{Fujitani82} was cited. \\

Our calculation involves the numerical integration, for which we use {the} software of
Wolfram Mathematica$^{\tiny \textregistered}$ ver.~10 (Wolfram Research).
Our formulation is stated in Sect.~\ref{sec:formulations}. 
We show the previous results in Sect.~\ref{sec:previousresults}.
Our results are shown in Sect.~\ref{sec:results}, and some details of the procedure are relegated to {Appendices}.  
The formulation and most part of the procedure are the same as given in \citet{Fujitani82};
we here show their key points indispensable for this paper to be self-contained.  The last section is devoted to
discussion.  

\section{Formulation \label{sec:formulations}}
\begin{figure}
 \begin{minipage}{0.45\hsize}
  \begin{center}
   \includegraphics[width=70mm
   ]{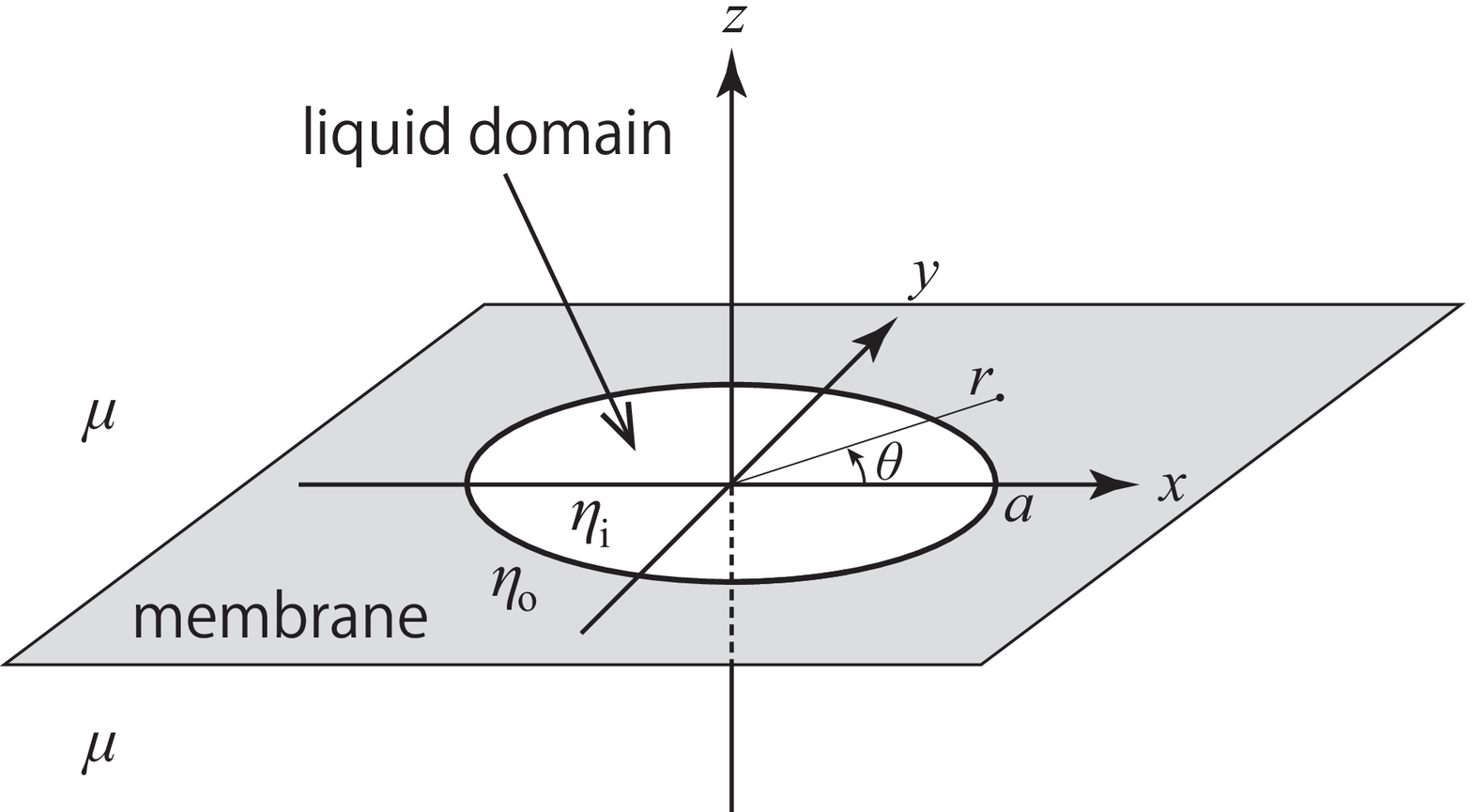}
  \end{center}
  \caption{
   A raft-like liquid domain with radius $a$ and viscosity $\eta_{\rm i}$ is embedded in a fluid membrane stretching on the $xy$-plane 
infinitely. The membrane viscosity outside the domain is denoted by $\eta_{\rm o}$.  The semi-infinite spaces on both sides of the membrane
are fulfilled by 3D fluids sharing the same viscosity $\mu$. 
}
  \label{fig1}
 \end{minipage}
 \hspace{0.05\hsize}
 \begin{minipage}{0.45\hsize}
  \begin{center}
   \includegraphics[width=65mm
   ]{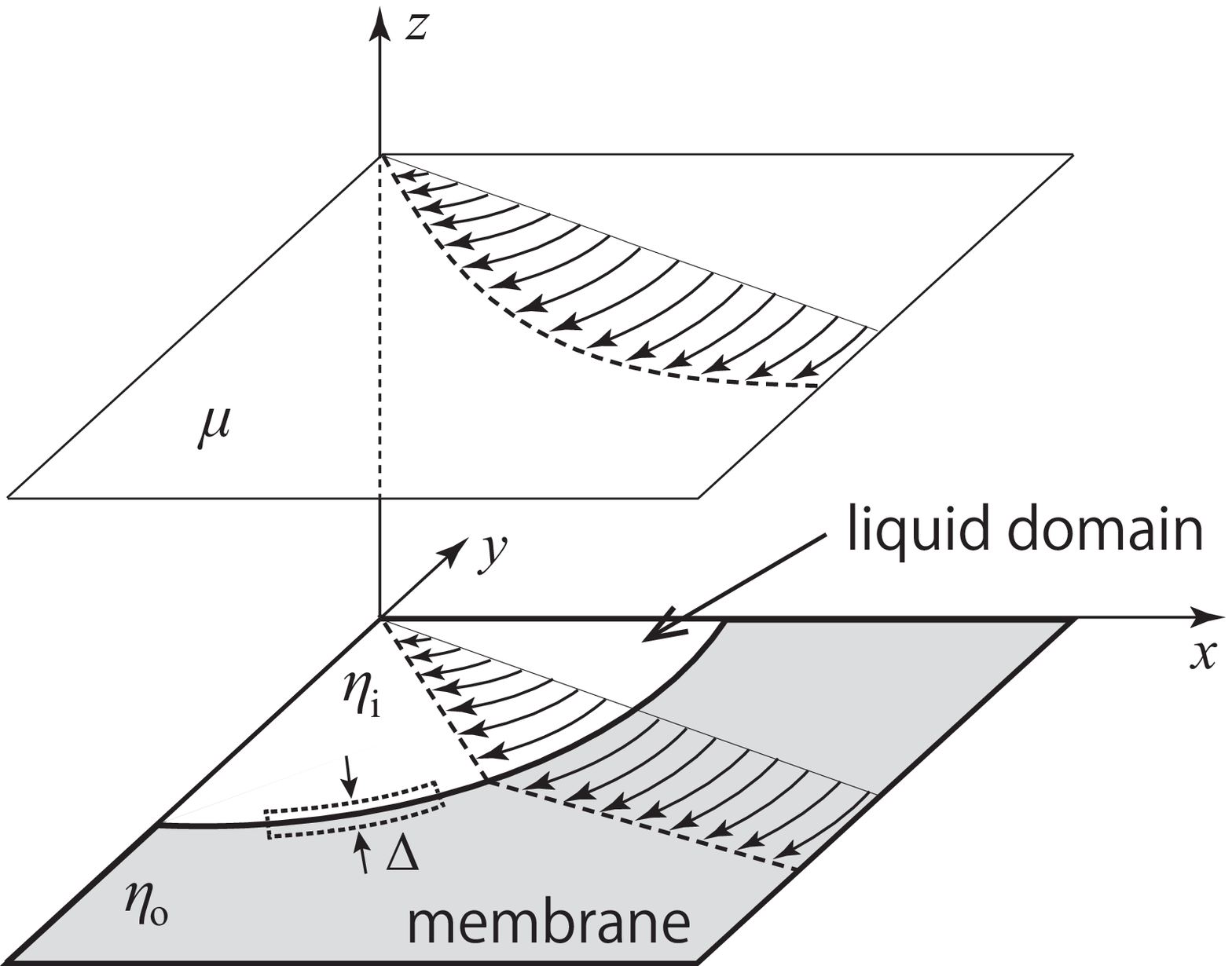}
  \end{center}
  \caption{ {A region of $x > 0$, $y < 0$ and $z\ge 0$ of fig.~\ref{fig1} is drawn. 
  The curved arrows represent velocity fields.
In the membrane, the velocity gradient should be discontinuous across the domain perimeter 
when $\eta_{\rm i}$ is not equal to $\eta_{\rm o}$.  In the 3D fluid, the velocity gradient
is continuous, as shown in
a cross section parallel to the membrane.   A slender region is drawn with the dotted closed curve
along the domain perimeter in the membrane.}
We write $\Delta$ for the size in the direction vertical to the perimeter. 
  }
  \label{fig2}
 \end{minipage}
\end{figure}

As is shown schematically in fig.~\ref{fig1}, 
the membrane is assumed to lie on the $xy$-plane of the Cartesian coordinate system $(x,y,z)$.
We also set the cylindrical coordinates $(r, \theta, z)$ so that the line $\theta=0$ is the $x$-axis.
A circular liquid domain with the radius $a$ shifts 
translationally with the velocity $\bmU =U \bme_x$, 
where $\bme_x$ denotes the unit vector of the $x$-axis.  We consider
the instant when the center coincides with the origin.  
The 3D fluids on both sides of the membrane share the same viscosity $\mu$.
The drag force can be written as ${\cal F}_x \,{\bme}_x$,
and the drag coefficient $\gamma$ is given by $-{\cal F}_x/U$.\\

The velocity field in the 3D fluids and that of the fluid membrane are respectively denoted by
${\bmV}$ and ${\bmv}$.  In this setting, we have
\be 
\lim_{r\to a} v_r (r,\theta)= U_r \ , \label{vrU}
\ee 
where $v_r$ and $U_r$ denote the $r$-components of ${\bmv}$ and ${\bmU}$, respectively.  
Because of the no-slip condition, we also have
\be
	\lim_{z \to 0} {\bmV}(r, \theta, z) = {\bmv}(r,\theta)\ . \label{noslip}
\ee
Suppose a 2D region along the domain perimeter, as shown in fig.~\ref{fig2}; we write
$\Delta$ for its width in the direction vertical to the perimeter.
As the perimeter becomes thin, or $\Delta$ approaches zero,
the forces exerted on the region should become balanced
and the stress exerted by the ambient 3D fluids becomes negligible.
Thus, the tangential stress of the 2D fluid should be continuous across the perimeter, which is represented by  
\be
	  \lim_{r\to a+}{ \tau}_{r\theta}(r,\theta)=
\lim_{r\to a-}{ \tau}_{r\theta}(r,\theta) \label{eqn:bctau}\ . 
\ee
Here, $\tau$ denotes the stress tensor associated with ${\bmv}$, 
and the limit $r \to a\pm$ indicates that $r$ approaches $a$ with $r-a$ being kept positive (negative).
When $\eta_{\rm i}$ equals $\eta_{\rm o}$, we can do without eq.~(\ref{eqn:bctau}) because
eq.~(\ref{eqn:bctau}) is automatically
satisfied when eq.~(\ref{noslip}) holds and ${\mbox{\boldmath $V$}}$ is smooth.
However, when $\eta_{\rm i}$ is not equal to $\eta_{\rm o}$,
eq.~(\ref{eqn:bctau})  makes the velocity gradient of the 2D fluid discontinuous across the perimeter, 
as shown in fig.~\ref{fig2}.  Then,
we should require both eqs.~(\ref{noslip}) and (\ref{eqn:bctau}).
This point is overlooked in \citet{Fujitani80}, as discussed later.  \\

The 3D velocity field satisfies
the Stokes equation and the incompressibility condition,
\be
	\mu \Delta {\bmV} = \nabla P  \quad {\rm and} \quad \nabla \cdot {\bmV} = 0\ , \label{eqn:3D}
\ee
where $P$ denotes the pressure field.
Equation~(\ref{eqn:3D}) holds for $z \neq 0$.
The 2D velocity field of the membrane fluid also satisfies  the Stokes equation and the incompressibility condition,
\be
	\eta \Delta {\bmv} + {\mbox{\boldmath $F$}} = \nabla p \quad {\rm and} \quad \nabla \cdot {\bmv} = 0\ , \label{eqn:2D}
\ee
where $p$ denotes the in-plane pressure field of the 2D fluid and ${\mbox{\boldmath $F$}}$ denotes the stress exerted by the 3D fluids.
The differential operators are defined in terms of $x$ and $y$ in eq.~(\ref{eqn:2D}), which holds for $z = 0$ and $r \ne a$.
The membrane viscosity $\eta$ equals $\eta_{\rm i}$ inside the domain $(r < a)$ and 
equals $\eta_{\rm o}$ outside the domain $(r >a)$. \\


\section{Previous results \label{sec:previousresults}}
We introduce the Fourier transforms with respect to $\theta$, {\it e.g.\/},
\be
	\tilde{V}_{zm}(r, z) = \frac{1}{2\pi} \int^{2\pi}_0 \d \theta \,\,V_z(r, \theta, z) \e^{-\smalli m \theta}
\label{tilVZM}\ee
with $m = 0, \pm 1, \pm2, \dots$, and the Hankel transforms with respect to $r$, {\it e.g.\/},
\be
	\hat{V}_{zm}(\zeta, z) = \int^\infty_0 \d r \,\, r J_m(\zeta r)\tilde{V}_{zm}(r, z)\ ,\label{tilVZM2}
\ee
where $J_m$ is the Bessel function of the first kind.
Because of the symmetry, only the fields with $m=\pm 1$ do not vanish.
In each field, the transforms of $m=\pm 1$ are related with each other.  
Thus, we have only to consider the fields with $m=1$.  
As shown in Appendix A, we rewrite
eq.~(\ref{eqn:3D}) into the Hankel transforms and solve the resultant equations for $m=1$
with two functions of $\zeta$ being left undetermined.  We can relate the two functions
with the aid of the second equation of eq.~(\ref{eqn:2D}), 
and thus have only to consider one undetermined function of $\zeta$.
We use 
\be
R \equiv {r\over a}\quad {\rm and}\quad \nu_{\rm o} \equiv {\eta_{\rm o}\over 2\mu a} \label{nu}
\ .\ee
Utilizing that the left-hand side (lhs) of eq.~(\ref{eqn:2D}) is irrotational, 
and calculating ${\mbox{\boldmath $F$}}$ in terms of ${\bmV}$, we {obtain}
\be
	0 &=& \int^\infty_0 \d \zeta \,\, \zeta^2 J_1(\zeta R)A(\zeta) \qquad {\rm for} \,\, 1 < R, \quad {\rm and} \label{dualint1}\\
	0 &=& \int^\infty_0 \d \zeta \,\, \zeta^2 J_1(\zeta R)A(\zeta) 
\left( 1 - \frac{\kappa}{1 + \nu_{\rm o} \, \zeta} \right) \qquad {\rm for} \,\, 0 \le R < 1\ , \label{dualint2}
\ee 
where $A$ is the undetermined function.    In \citet{Fujitani82},
the 3D fluid on each side of the membrane 
is assumed to be confined by the membrane and a wall, which is parallel to the membrane and lies 
at the distance $H$ from the membrane.   Taking the limit of $H \to \infty$ in
eqs.~(3.1) and (3.2) in \citet{Fujitani82} 
gives eqs.~(\ref{dualint1}) and (\ref{dualint2}) above.
Equation (\ref{vrU}) yields
\be
  2\mu a^2 U= \int_{0}^\infty \d \zeta {J_1(\zeta)A(\zeta)\over \zeta\left(1+\nu_{\rm o}\zeta\right)} \label{eqn:vrU2},
\ee 
which corresponds with {eq.~(2.37) of \citet{Fujitani82} because of eqs.~(2.26), (3.8)}, and (3.13) there. \\

Let us define $q(R)$ as  the integral of eq.~(\ref{dualint1}) for $R\ge 0$.
As shown in \citet{Fujitani82}, we can arrive at the solution by assuming
\be
  q(R)=q_1(R)+ c_1 \delta\left(R-1\right) +c_2 {{\rm d}\over {\rm d}R} \delta\left(R-1\right)
\ ,\label{eqn:qR}
\ee
where $q_1(R)$ is a finite function vanishing for $R>1$, and $c_1$ and $c_2$ are constants independent of $R$. 
The third term on the right-hand side (rhs) above is missed in \citet{Fujitani80}.
Without this term, we can satisfy all the conditions other than eq.~(\ref{eqn:bctau}), but 
{the resultant} solution is naturally incorrect unless $\eta_{\rm i}$ equals $\eta_{\rm o}$.  
The third term is taken into account in \citet{Fujitani82}.
The second term gives a point source, while the third term gives a 
dipole source, which is analogous to the single-layer and double-layer potentials in the boundary integral,
respectively \citep{Pozrikidis}. 
The Hankel transformation of eq.~(\ref{eqn:qR}) involves the integral $Rq_1(R)J_1(\zeta R)$ over $0<R<1$.
Rewriting this integral with the aid of eq.~(\ref{dualint2}), we obtain
\be
&&\zeta A(\zeta)=c_1 J_1(\zeta)-c_2\zeta J_0(\zeta) +\kappa \int_0^\infty \d \xi \ {\xi^2A(\xi)\over 1+\nu_{\rm o}\xi}
  \int_0^1 \d R\ RJ_1(\zeta R)J_1(\xi R) \label{eqn:azeta}\\
&&\qquad =c_1 J_1(\zeta)-c_2\zeta J_0(\zeta) +\kappa \, \zeta \int_0^\infty \d \xi \ K(\zeta,\xi) A(\xi)
\ , \label{eqn:Azeta}
\ee
where the kernel is defined as
\be
K(\zeta, \xi)\equiv \frac{\xi^2}{\left(1+ \nu_{\rm o}\xi\right)\zeta} \times \frac{\xi J_0(\xi)J_1(\zeta)-\zeta J_0(\zeta)
J_1(\xi)}{\zeta^2-{\xi}^2}
\ .\label{truekernel}
\ee
Equation (\ref{eqn:azeta}) is essentially derived in Sect.~3.2 of \citet{Fujitani82}
although not shown explicitly because the discussion is mainly focused on
the order of $\kappa$ there. 
The constants $c_1$ and $c_2$ can be determined with the aid of eqs.~(\ref{eqn:bctau}) and (\ref{eqn:vrU2}).  
Another kernel, $M$, is used in \citet{Fujitani82}; its definition and relation to $K$ are given by
\be
	M(\zeta, \xi) &\equiv& 
\frac{\xi}{1+ \nu_{\rm o}\xi} \times \frac{\zeta J_0(\xi)J_1(\zeta)-\xi J_0(\zeta)
J_1(\xi)}{\zeta^2-{\xi}^2} \label{kernelM}\\
&=& K(\zeta, \xi)+ {\xi J_0(\xi)J_1(\zeta)\over \left(1+\nu_{\rm o}\xi\right)\zeta}
\ .\label{kernel}
\ee
Calculating the total force exerted on the liquid domain, as shown in \citet{Fujitani82},
we find the drag coefficient to be given by
\be
&& \gamma = {\pi\over aU}\lim_{R\to 1+}\lim_{Z\to 0+}
\int_0^\infty\d \zeta\ \zeta J_2(\zeta R)
A(\zeta)\e^{-\zeta Z} \ ,\label{gam}
\ee
where we use $Z\equiv z/a$.  
This equation is easily derived from eqs.~(2.24), (2.41) and (3.13) of \citet{Fujitani82}.\\

When $\kappa$ vanishes, we have $c_2 = 0$ and find
\be
A(\zeta)={2\mu a^2 U\over Y_0(\nu_{\rm o})}{J_1(\zeta)\over \zeta}\quad {\rm for}\ \kappa=0 
\ ,\label{kok}
\ee
which is substituted into eq.~(\ref{gam}) to yield the previous result of \citet{DeKoker}.  
 Writing $\gamma_{\kappa=0}$ for his result of the
drag coefficient, we have
\be
	\gamma_{\kappa=0} = {2\pi\mu a \over Y_0(\nu_{\rm o})} 
	\label{gammakok}
\ ,\ee
where $Y_0$ is defined as 
\be
	Y_0(\nu_{\rm o}) = \int_0^\infty \d \zeta\ \frac{J_1^2(\zeta)}{\zeta^2(1 + \nu_{\rm o} \zeta)}\ . \label{Y0}
\ee

For a disk, eq.~(\ref{vrU}) is replaced by ${\bmv} = {\bmU}$ for 
$r \le a$, and eq.~(\ref{eqn:bctau}) is not required, as was {discussed} in \citet{Saffman} and \citet{Hughes}.
Equation (3.50) in the latter reference can be rewritten as a set of simultaneous equations with respect to
$\omega_0, \omega_1,\ldots$ represented by
\be
	\delta_{l, 0} 
= \sum^\infty_{m = 0} \omega_m \left[ \int^\infty_0 {\d u} \, \frac{ j_{2l}(u) j_{2m}(u)}{u + \nu_{\rm o}^{-1}} \right]
\quad {\rm for}\ l=0,1,2,\ldots\ . \label{Lambda}
\ee
Here, $\delta_{i, j}$ denotes Kronecker's delta and
$j_{0}(u), j_2(u), \ldots$ denote the spherical Bessel functions.
According to \citet{Hughes}, $\omega_0$ is related with the drag coefficient of a disk, 
for which we write $\gamma_{\rm disk}$, as
\be
\gamma_{\rm disk}=4\pi\eta_{\rm o} \omega_0 \label{gammadisk}
\ee 
Because $\omega_0$ is determined only by $\nu_{\rm o}$, it is convenient to introduce a dimensionless
drag coefficient,
\be
	\gamma^*_{\rm disk}(\nu_{\rm o}) \equiv \frac{\gamma_{\rm disk}}{4\pi \eta_{\rm o}} \ . \label{gammastar}
\ee 
Similarly, we find the quotient of eq.~(\ref{gammakok}) divided by $4\pi\eta_{\rm o}$ to be
determined only by $\nu_{\rm o}$.  We write $\gamma^*(\nu_{\rm o}, 0)$ for the quotient;
the zero in the parentheses means $\kappa=0$.  Equation (\ref{gammakok}) gives
\be
\gamma^*(\nu_{\rm o}, 0)= {1\over 4\nu_{\rm o}Y_0(\nu_{\rm o})}\ .
\ee
The ratio of $\gamma_{\rm disk}$ to $\gamma_{\kappa=0}$ equals  
that of $\gamma^*_{\rm disk}(\nu_{\rm o})$ to $\gamma^*(\nu_{\rm o},0)$, 
and thus depends only on $\nu_{\rm o}$.  We write 
${\tilde \gamma}_{\rm disk}(\nu_{\rm o})$ for the ratio, {\it i.e.\/},
\begin{equation}
{\tilde \gamma}_{\rm disk}(\nu_{\rm o})={\gamma_{\rm disk} \over \gamma_{\kappa=0}}
={\gamma^*_{\rm disk}(\nu_{\rm o}) \over\gamma^*(\nu_{\rm o},0)}
\end{equation}
 Typical values of $\mu$ and $\eta_{\rm o}$ are respectively $1 \,\,\,{\rm g/(m \, s)}$ and $10^{-7} 
\,\,\, {\rm g/s}$ \citep{Merkel, Smeulders}.  
We can calculate ${\gamma}^*_{\rm disk}(\nu_{\rm o})$ by   
truncating the sum in eq.~(\ref{Lambda}) up to $m = 20$; the absolute value of the
change in the result obtained  when we use the sum up to $m = 21$ 
is much smaller than $10^{-3}$
for each of the values $\nu_{\rm o}$ considered.
The previous results for $\nu_{\rm o} = 10, 1$ and $0.1$ are
summarized in Table~\ref{tab1}.

\begin{table}
  \begin{center}
  \begin{tabular}{ccccc}
     $\nu_{\rm o}$ & \,\,$a \,\, [{\rm nm}]^\dagger$\,\, & \,$ \,\, 
\gamma^*_{\rm disk}(\nu_{\rm o})/\gamma^*(\nu_{\rm o}, 0)={\tilde\gamma}_{\rm disk}(\nu_{\rm o})$ 
\,\, & \,$\gamma_{\rm disk}\ [\mu {\rm g}/{\rm s}]^\dagger /\gamma_{\kappa=0} \ [\mu {\rm g}/{\rm s}]^\dagger=
{\tilde\gamma}_{\rm disk}(\nu_{\rm o}) \,\, $  \\[0.5em] 
    10 & 5  &  $0.3952/0.3635= 1.087$ & $0.4966 /0.4568 = 1.087 $  \\[0.3em] 
    1 & 50 & $1.212/1.067=1.137$ & $1.524 /1.340 =1.137$ \\[0.3em] 
    0.1 & 500 & $7.317/6.513=1.123$ & $9.194/8.184 = 1.123$ \\[0.3em]
  \end{tabular}
  \caption{Previous results. \newline
$\dagger$: Values calculated by using typical values of $\mu$ and $\eta_{\rm o}$ mentioned in the text.}
  \label{tab1}
  \end{center}
\end{table}


\section{Results \label{sec:results}}
\subsection{Analytical results}
Considering eq.~(\ref{kok}), it is convenient to introduce
\be 
	A(\zeta) = {2\mu a^2 U\over Y_0(\nu_{\rm o})} \tilde{A}(\zeta)\ .\label{eqn:defA}
\ee
We expand ${\tilde A}$ with respect to $\kappa$ as
\be
	\tilde{A}(\zeta) &=& \sum_{n = 0}^\infty \kappa^n {\tilde A}_n(\zeta)\ , \label{tilan}
\ee
where 
\be
{\tilde A}_0(\zeta)={J_1(\zeta)\over\zeta}\ .\label{tilanzero}
\ee
Let us define an operator ${\hat M}$ as
\be
    \left[\hat{M}f\right](\zeta) \equiv \int^\infty_0 \d \xi M(\zeta, \xi) \, f(\xi)\ ,\label{M}
    \ee
where $f$ is a function.    From eq.~(\ref{eqn:Azeta}), as shown in Appendix B, we derive 
\be
{\tilde A}_n(\zeta) &=& \alpha_n \frac{J_1(\zeta)}{\zeta} + \beta_n J_0(\zeta) + 
 \left[{\hat M}{\tilde A}_{n-1}\right](\zeta) \quad {\rm for}\ n=0,1,2,\ldots \ .\label{tilan2}
 \ee
Here, we stipulate $\alpha_0=1$, $\beta_0=0$ and ${\tilde A}_{-1}=0$ because of eq.~(\ref{tilanzero}).
The constants $\alpha_n$ and $\beta_n$ for $n\ge 1$ depend only on $\nu_{\rm o}$, as shown below.  
\\

Substituting eqs.~(\ref{eqn:defA}), (\ref{tilan}), and (\ref{tilan2}) into eq.~(\ref{eqn:vrU2}) yields 
\be
    \alpha_n = - \frac{1}{Y_0} \left( \beta_n X + \hat{L}\hat{M}{\tilde A}_{n - 1} \right) 
\quad {\rm for} \ n=1,2,\ldots
\ .\label{alpha} 
\ee    
As shown in Appendix B, substituting eqs.~(\ref{eqn:defA}), 
(\ref{tilan}), and (\ref{tilan2}) into eq.~(\ref{eqn:bctau}) yields
\be
	\beta_n = \nu_{\rm o} \left( \alpha_{n - 1} {\hat N}{\tilde A}_0 +  \beta_{n - 1} {\cal G}+\hat{N} \hat{M} {\tilde A}_{n - 2} 
      \right) \quad {\rm for} \ n=1,2,\ldots \ . \label{beta}
\ee 
Here, we use                  
\be
	&&X \equiv \int_0^\infty \d \zeta \frac{J_1(\zeta) J_0(\zeta)}{\zeta(1 + \nu_{\rm o} \zeta)}\ ,\quad
	{\cal G} \equiv 
\lim_{R \to 1+} \int^\infty_0 \d \zeta \frac{ \zeta J^\prime_2 (\zeta R) J_0(\zeta)}{1 + \nu_{\rm o} \zeta} \label{G+} \ ,\\
	&& 
\hat{N}f \equiv \int^\infty_0 \d \zeta \frac{\zeta J^\prime_2(\zeta)}{1 + \nu_{\rm o} \zeta} \, f(\zeta)\ ,  \,\,\,{\rm and}
\,\, 
\hat{L}f \equiv \int^\infty_0 \d \zeta \frac{J_1(\zeta)}{\zeta(1 + \nu_{\rm o} \zeta)} \, f(\zeta) \, ,  \label{Lf}
\ee 
where $f$ is a function and $J'_2(\zeta)$ implies
\begin{equation}
J'_2(\zeta)\equiv {{\rm d} J_2(\zeta)\over{\rm d}\zeta}=J_1(\zeta)-{2J_2(\zeta)\over \zeta}
\ .\end{equation} 
We have $Y_0={\hat L}{\tilde A}_0$.  The integral in the second equation of eq.~(\ref{G+}) is
a function of $R$ and is discontinuous at $R=1$ because the integrand does not approach zero so rapidly.
\\

In general, the drag coefficient $\gamma$ depends on $a$, $\mu$, $\eta_{\rm o}$, and
$\eta_{\rm i}$.
In eq.~(\ref{gammakok}), ${\gamma}_{\kappa=0}$ depends on $a$, $\mu$ and $\eta_{\rm o}$.
As shown below, the ratio $\gamma/ {\gamma}_{\kappa=0}$ depends only on $\nu_{\rm o}$
and $\kappa$.
Substituting eqs.~(\ref{eqn:defA}), (\ref{tilan}), and (\ref{tilan2}) into eq.~(\ref{gam}), we obtain
\be
	\gamma
&=& \gamma_{\kappa=0} \,
\tilde{\gamma}(\nu_{\rm o}, \kappa) \ ,  \label{gamma}
\ee
where ${\tilde \gamma}$ is defined as 
\be
	\tilde{\gamma}(\nu_{\rm o}, \kappa) =\sum_{n = 0}^\infty \, \tilde{\gamma}_n(\nu_{\rm o})
 \, \kappa^n \ , \label{tilgamma}
\ee
with ${\tilde \gamma}_n$ being given by 
\be
	\tilde{\gamma}_n(\nu_{\rm o}) =\alpha_n + 2\beta_n + \lim_{R \to 1+}\, 
\left[ \int_0^\infty \d \zeta \,\, \zeta J_2(\zeta R) \left[{\hat M}{\tilde A}_{n - 1}\right] (\zeta) \right] 
\quad {\rm for}\ n=1,2,\ldots \label{gamman}
\ee
and  $\tilde{\gamma}_0(\nu_{\rm o})  = 1$.  This equality implies ${\tilde \gamma}(\nu_{\rm o},0)=1$, as it should be. 
The term in the braces of eq.~(\ref{gamman}) converges without the factor
$\e^{-\zeta Z}$ appearing in eq.~(\ref{gam}) but is discontinuous at $R=1$.



\subsection{Numerical results}
\begin{figure}
  \begin{center}
  \includegraphics[angle=0,width=9.0cm
  ]{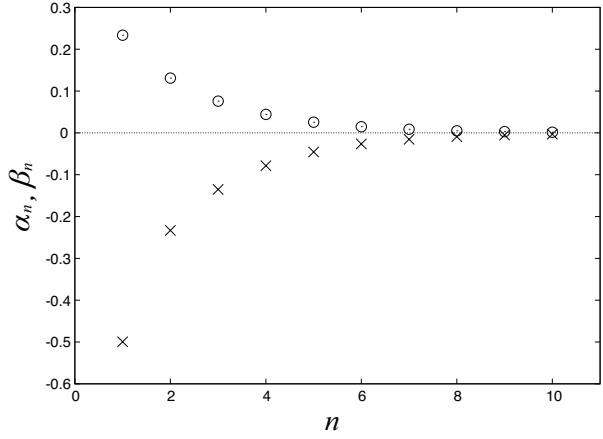}
  \caption{Numerical results of $\alpha_n$ (circles) and $\beta_n$ (crosses) for $\nu_{\rm o} = 10$. 
  }
  \label{fig-alphabeta}
  \end{center}
\end{figure}

To obtain $\alpha_n$ and $\beta_n$ successively by using eqs.~(\ref{alpha}) and (\ref{beta}),
we calculate the integrals appearing in eqs.~(\ref{Y0}), (\ref{M}), (\ref{G+}) and (\ref{Lf}).
Among them, we should replace the upper bounds $\infty$ in eqs.~(\ref{M}) and (\ref{Lf}) with finite values for numerical calculation;
both of the values are determined to be $10000$ so that the integrals appear to remain unchanged even if the
upper bounds are made to be larger. 
The other integrals can be computed simply, except for the second equation of 
eq.~(\ref{G+}); it becomes a continuous function of $R$ changing rapidly across $R = 1$.
Thus, we should estimate the value at $R\to 1+$ by using the results of numerical integration
at values of $R$ larger than and close to the unity.  
We estimate ${\cal G}$ by using the value at $R=1.005$ for $\nu_{\rm o} = 10$, 
$1.0001$ for $\nu_{\rm o} = 1$, and $1.00001$ for $\nu_{\rm o} = 0.1$, respectively.
The results for $\nu_{\rm o} = 10$ are shown in fig.~\ref{fig-alphabeta}, where
$\alpha_n$ and $\left|\beta_n\right|$ decrease monotonically as $n$ increases.
They approach zero more slowly for
$\nu_{\rm o} = 1$ and $\nu_{\rm o} = 0.1$ although data not shown.  \\

\begin{figure}
  \begin{center}
  \includegraphics[angle=0,width=10.0cm
  ]{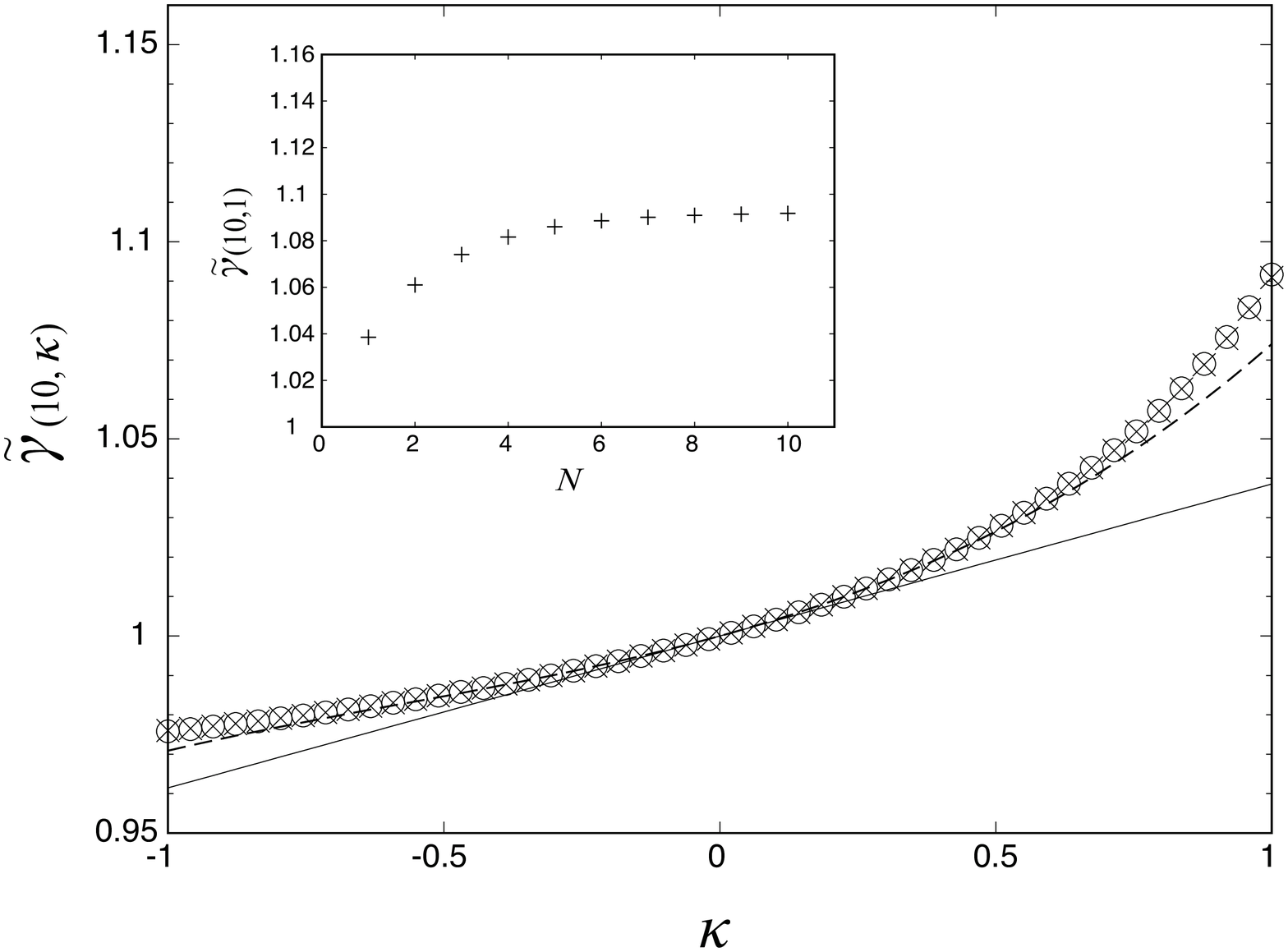}
  \caption{Numerical results of eq.~(\ref{truncatedgamma}) for $\nu_{\rm o} = 10$ 
are shown for $N = 1$ (solid line), $3$ (dotted line), $8$ (crosses) and $10$ (circles).
		The inset shows how the numerical results at $\kappa=1$ depend on $N$.}
  \label{fig-nu10}
  \end{center}
\end{figure}
\begin{figure}
  \begin{center}
  \includegraphics[angle=0,width=10.0cm
  ]{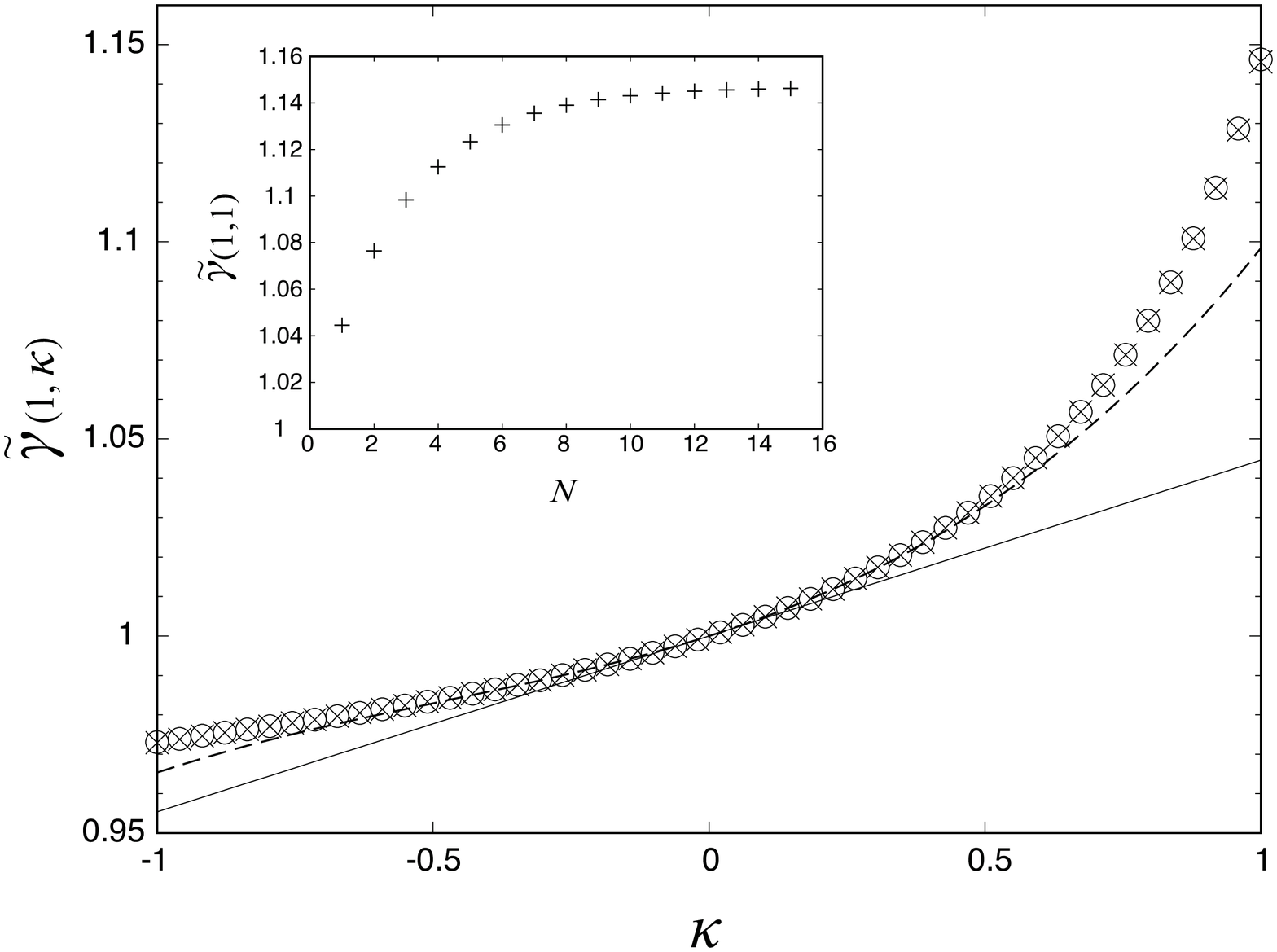}
  \caption{Numerical results of eq.~(\ref{truncatedgamma}) for $\nu_{\rm o} = 1$ are shown for $N = 1$ (solid line), $3$ (dotted line), $12$ (crosses) and $15$ (circles).
		The inset shows how the numerical results at $\kappa=1$ depend on $N$.
  }
  \label{fig-nu1}
  \end{center}
\end{figure}
\begin{figure}
  \begin{center}
  \includegraphics[angle=0,width=10.0cm
  ]{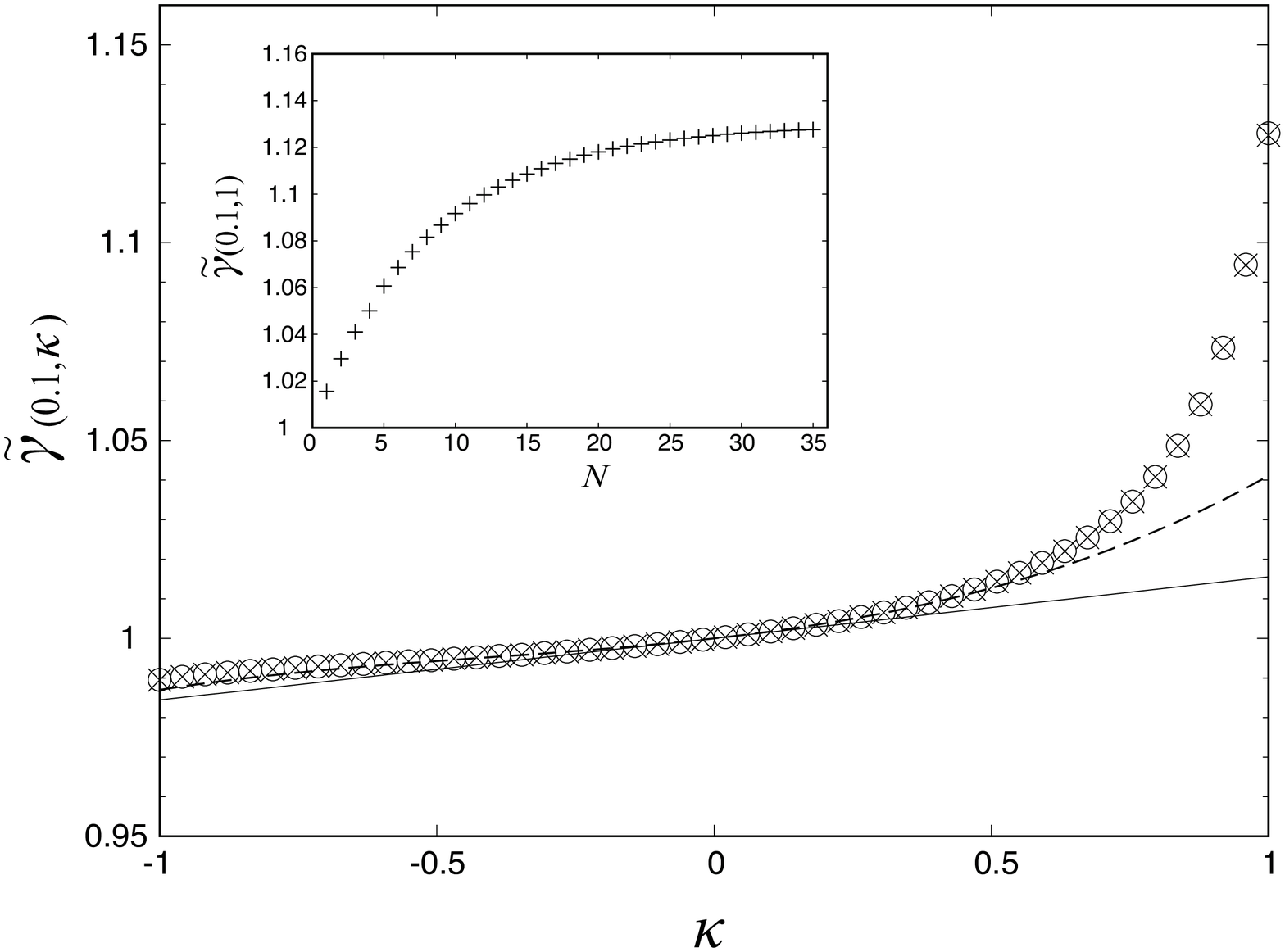}
  \caption{Numerical results of eq.~(\ref{truncatedgamma}) for $\nu_{\rm o} = 0.1$ are shown for $N = 1$ (solid line), $3$ (dotted line), $33$ (crosses) and $35$ (circles).
		The inset shows how the numerical results at $\kappa=1$ depend on $N$.
  }
  \label{fig-nu01}
  \end{center}
\end{figure}
%


In calculating  $[\hat{M}\tilde{A}_{n-1}]$ of eq.~(\ref{gamman}), 
we encounter the integral $\hat{M} J_0$, which converges 
 but needs larger upper bound for numerical calculation.
However, judging from our numerical results not shown here,
we can expect well that this integral is much smaller than that of the other terms in $[\hat{M}\tilde{A}_{n-1}]$. 
Thus, we estimate $\hat{M} J_0$ by setting the upper bound to be $10000$. 
In estimating the value of the integral appearing explicitly in eq.~(\ref{gamman}), 
we set the option of Mathematica$^{\tiny \textregistered}$ as Global Adaptive and choose the Max Error Increases as $10000$.
The results for each of $\nu_{\rm o}=10, 1,$ and $0.1$ represent a function of $R$, which 
oscillates near $R = 1$ probably because the upper bounds of the integrals are changed to finite values.
Instead of raising the upper bound of the integral from $10000$, 
we estimate the value in the limit of $R \to 1+$ by extrapolating the least-squares linear-regression equation
which we obtain by using the numerical results 
from $R = 1.001$ to $1.01$ with the intervals being $0.0001$.
 \\

For numerical calculation of eq.~(\ref{tilgamma}), we should
truncate the series into the sum of the first $N$ terms, {\it i.e.}, 
\be
\sum_{n = 0}^N \, \tilde{\gamma}_n(\nu_{\rm o}) \kappa^n
\ . \label{truncatedgamma}
\ee
Numerical results for various values of $N$ are shown for different values of $\nu_{\rm o}$ in
figs.~\ref{fig-nu10}, \ref{fig-nu1}, and \ref{fig-nu01}.
The results for $N=1$ are also calculated in \citet{Fujitani82}.
In each of the figures, eq.~(\ref{truncatedgamma}) increases with $N$ for any nonzero $\kappa$.
As shown by each inset figure, the increment of eq.~(\ref{truncatedgamma}) occurring  at $\kappa=1$
when $N$ increases by one decreases with $N$.
For $\nu_{\rm o}=10$, the increment is much smaller than $10^{-3}$ for $N \ge 9$, and   
we regard eq.~(\ref{truncatedgamma}) with $N=10$ as
the value of eq.~(\ref{tilgamma}).   
For $\nu_{\rm o}=1$, the increment is much smaller than $10^{-3}$ for $N \ge 14$,
and we regard eq.~(\ref{truncatedgamma}) with  $N = 15$ as the value
of eq.~(\ref{tilgamma}).   
For $\nu_{\rm o}=0.1$, the increment is much smaller than $10^{-3}$ for $N \ge 34$, 
and we regard eq.~(\ref{truncatedgamma}) with $N = 35$ as the value of eq.~(\ref{tilgamma}).

%

%


\section{Discussion}

We study the drag coefficient $\gamma$ of a liquid domain in a fluid membrane immersed in a 3D
fluid.  We extend 
the previous calculation up to the order of $\kappa$ in \citet{Fujitani82} to derive
the recursion relations of the coefficients appearing in the series expansion of ${\tilde \gamma}(\nu_{\rm o}, \kappa)$ 
with respect to $\kappa$,
as shown by eqs.~(\ref{alpha}), (\ref{beta}), and (\ref{gamman}).  See eqs.~(\ref{gammakok}) and (\ref{gamma}) for 
the relation between $\gamma$ and ${\tilde \gamma}$.
We numerically examine how the partial sum of eq.~(\ref{truncatedgamma}) depends on $N$, and
find that the sums for $N=10$, $15$, and $35$ 
can be identified with eq.~(\ref{tilgamma}) for $\nu_{\rm o}=10$, $1$, and $0.1$, respectively. \\

The slopes of the solid lines in figs.~\ref{fig-nu10}-\ref{fig-nu01}
are respectively $0.0385$,  $0.0446$, and $0.0156$, which agree well with 
the corresponding values in the column of $H\to \infty$ of Table II of \citet{Fujitani82}.
For each of $\nu_{\rm o}=10$ and $1$, 
the sum up to $N=1$
gives a good approximation of ${\tilde \gamma}$ when $|\kappa|$ is smaller than about $0.2$.
For $\nu_{\rm o} = 1$, the region of $\kappa$ where the sum up to $N=1$ is available is 
wider. 
When $\kappa$ is closer to unity beyond the region in each of the figures, the derivative of ${\tilde \gamma}$
with respect to $\kappa$ becomes larger.  
The drag coefficient is expected to reach a plateau value depending on 
the parameter as the ratio $\eta_{\rm i}/\eta_{\rm o}$ approaches zero. The circles for $\kappa < 0$ 
in any of figs.~\ref{fig-nu10}-\ref{fig-nu01}  appears to be consistent with this expectation. 
It remains to be studied whether the series (\ref{tilgamma})
is convergent or asymptotic. If the series has the radius of convergence, it would be unity, 
considering that the value of $\kappa$ larger than unity is meaningless. 
The dependence of the drag coefficient on $\kappa$ for $\kappa>0$ is not pointed out in fig.~6a of \citet{RD}; their $\beta$ and $\epsilon$
are respectively our $\nu_{\rm o}$ and $\kappa$.
This figure is obtained from their eq.~(2.7); this integral equation is discretized into a set of 
some thousands of simultaneous equations for numerical calculation. 
In the present study, to derive an equation corresponding with their integral equation above,
we can sum up eq.~(\ref{tilan2}) from $n=0$ to $\infty$ 
and use eq.~(\ref{tilan}) to derive an integral equation with respect to ${\tilde A}(\zeta)$. 
This equation contains a term involving $J_0$,  unlike 
eq.~(2.7) of \citet{RD}.
This is because the third term on the lhs
of eq.~(\ref{eqn:qR}) is not considered in \citet{RD}.  This term is required unless $\eta_{\rm i}$ equals $\eta_{\rm o}$,
as discussed below eq.~(\ref{eqn:qR}). \\

\begin{table}
  \begin{center}
  \begin{tabular}{cccc}
    $\nu_{\rm o}$ & \,$\tilde{\gamma}(\nu_{\rm o}, 1)$\, & \,$\tilde{\gamma}_{\rm disk}(\nu_{\rm o})$ \\[0.5em] 
    10 & 1.092 & 1.087    \\[0.3em] 
    1 & 1.146 & 1.137    \\[0.3em] 
    0.1 & 1.128 & 1.123 \\[0.3em]
  \end{tabular}
  \caption{Comparison of our results at $\kappa=1$ with the previous results for a disk.
  }
  \label{tab2}
  \end{center}
\end{table}

Let us concentrate on our results at $\kappa = 1$; 
we have ${\tilde \gamma}(\nu_{\rm o}, 1)=1.092$, $1.146$, and $1.128$ for $\nu_{\rm o}=10$, $1$, and $0.1$, respectively,
as listed in Table~\ref{tab2}.   
These values are respectively in good agreement with the corresponding 
previous results of
$\tilde{\gamma}_{\rm disk}(\nu_{\rm o})$, 
which are also shown in Table \ref{tab1}.
This shows that the drag coefficient of a liquid domain 
reasonably approaches the one for a disk as $\eta_{\rm i}/\eta_{\rm o}$ tends to infinity.  
This strongly suggests that the procedure shown here is appropriate for calculating
the drag coefficient of a liquid domain.  \\

Using the procedure shown here, we can also calculate the drag coefficient of a liquid domain
embedded in a fluid membrane surrounded by 
confined 3D fluids; this case was considered 
only up to the order of $\kappa$ in \citet{Fujitani82}.
The procedure can be also applied in calculating  small deformation of a liquid domain
in a 2D linear shear flow; it was calculated in \citet{Fujitani74} only for
$\kappa=0$ and the stagnation flow.


\section*{Acknowledgements}
H. T. was financially supported by Organization for the Strategic Coordination of Research and Intellectual Properties in Meiji University.
Part of the work by Y. F. was financially supported by Keio Gakuji Shinko Shikin.

\section*{Appendix A: Derivation of eqs.~(\ref{dualint1}) and (\ref{dualint2})}  \label{subsec:AppA}
\renewcommand{\theequation}{A.\arabic{equation}}
\setcounter{equation}{0}


We introduce
$\tilde{V}_m^\pm = \tilde{V}_{r m} \pm {\rm i}\tilde{V}_{\theta m}$ and define $\hat{V}_m^\pm$ as
\be
	\hat{V}_m^\pm(\zeta, z) &=& \int^\infty_0 \d r \,\, rJ_{m\pm 1}(\zeta r) \tilde{V}_m^\pm(r,z)\ .
\ee
Using the transformations of eqs.~(\ref{tilVZM}) and (\ref{tilVZM2}), we can rewrite
eq.~(\ref{eqn:3D}) into
\be
	\mu(\partial^2_z - \zeta^2)\hat{V}_m^\pm &=& \mp \zeta \hat{P}_m\ , \label{3D-1}\\[0.3em]
	\mu(\partial^2_z - \zeta^2)\hat{V}_{z m} &=&\partial_z \hat{P}_m\ ,  \label{3D-2} \\[0.3em]
	 {\rm and} \quad \zeta \hat{V}^+_m - \zeta \hat{V}^-_m &=& -2\partial_z \hat{V}_{z m}\ . \label{3D-3}
\ee
As shown in \citet{Fujitani80}, we can solve eqs.~(\ref{3D-1})-(\ref{3D-3}) together with boundary conditions,
eq.~(\ref{vrU}) and (\ref{noslip}).
In particular, we have
\be
	\hat{P}_{\pm 1} &=& \frac{\zeta}{2}\left[ d^-_{\pm 1}(\zeta) - d^+_{\pm 1}(\zeta) \right]\e^{- \zeta |z|}\ , 
\ee
where $d_1^\pm$ and $d_{-1}^\pm$ are the coefficients to be determined.
From the symmetry, we have ${\hat P}_1 = -{\hat P}_{-1}$, which leads $d_1^--d_1^+ = d_{-1}^+-d_{-1}^-$.
Similarly, the symmetry of the velocity field yields $d_1^+ = d_{-1}^-$ and $d_1^- = d_{-1}^+$.
As discussed in \cite{Fujitani80} and \cite{Fujitani82}, we find $d_1^+ = d_1^-$ with the aid of
 the second equations of eqs.~(\ref{eqn:3D}) and (\ref{eqn:2D}). 
Thus, we arrive at 
\be
	\hat{P}_m = \hat{V}_{z m} = 0 \quad {\rm and}\quad 
	\hat{V}^\pm_m(\zeta) = \frac{\e^{- \zeta |z|}}{2\mu} d^\pm_m(\zeta) \quad {\rm for} \,\, m = \pm 1 \ .\label{Vz1}
\ee 
%
Since the lhs of eq.~(\ref{eqn:2D}) should be irrotational, we can eliminate $p$ from eq.~(\ref{eqn:2D}) to derive
\be
	0 = \int^\infty_0 \d \zeta \,\, \zeta^3 \, d^+_1(\zeta) \left( 1 + \frac{\eta}{2\mu} \zeta \right) J_1(\zeta r)\ , \label{2.20} 
\ee
for $0 \le r < a$ and $a < r$.   In these regions, respectively, $\eta$ equals $\eta_{\rm i}$ and $\eta_{\rm o}$.
Equation (\ref{2.20}) yields eqs.~(\ref{dualint1}) and (\ref{dualint2}) with $A(\zeta)$ being defined as 
\be
	A(\zeta) = 2\zeta(1 + \nu_{\rm o} \zeta) \, d_1^+\left({\zeta\over a}\right)\ . 
\ee

\section*{Appendix B: Derivation of the recursion equations}  \label{subsec:AppB}
\renewcommand{\theequation}{B.\arabic{equation}}
\setcounter{equation}{0}

We expand $c_1$ and $c_2$ in eq.~(\ref{eqn:Azeta}) with respect to $\kappa$ as
\begin{equation}
c_1={2\mu a^2U\over Y_0(\nu_{\rm o})}\sum_{n=0}^\infty \alpha_n^\sharp \kappa^n\quad {\rm and}
\quad c_2={2\mu a^2U\over Y_0(\nu_{\rm o})}\sum_{n=0}^\infty \beta_n \kappa^n\ ,
\end{equation}
where $\alpha^\sharp_n$ and $\beta_n$ are the expansion coefficients independent of $\kappa$.
Because of eq.~(\ref{kok}), we have $\alpha^\sharp_0=1$.
Substituting eqs.~(\ref{eqn:defA}) and (\ref{tilan}) into eq.~(\ref{eqn:Azeta}) yields eq.~(\ref{tilan2}), where
$\alpha_n$ is given by
\be
\alpha_n &=& \alpha^\sharp_n-\int^\infty_0 {\rm d} \xi \frac{\xi J_0(\xi)}{1 + \nu_{\rm o} \xi} {\tilde A}_{n-1}(\xi) \ .
\ee

\medskip
The Fourier transform of $\tau_{r \theta}$ is defined in the same way as in eq.~(\ref{tilVZM}).  For $m=1$, we have  
\be
	\tilde{\tau}_{r \theta 1} = 
\frac{- \i \eta}{4 \mu a^3} \int^\infty_0 \d \zeta \, \frac{\zeta J_2^\prime(\zeta R)A(\zeta)}{1 + \nu_{\rm o}\zeta}\ , \label{tiltau}
\ee
which comes from eqs.~(2.24), (2.41), and (3.13) of \cite{Fujitani82}.
Substituting eqs.~(\ref{eqn:defA}), (\ref{tilan}), and (\ref{tilan2}) into eq.~(\ref{tiltau}), 
we use eq.~(3.28) of \cite{Fujitani82} to find that eq.~(\ref{eqn:bctau}) gives
\be
\sum_{n=0}^\infty	\beta_n \left(-\kappa {\cal G} +\frac{1}{\nu_{\rm o}} \right) \kappa^n
= \sum_{n=0}^\infty \left( \alpha_n {\hat N}{\tilde A}_0+ \kappa  {\hat N}{\hat M} {\tilde A_n} \right)\kappa^{n+1}
\ , \label{bctau2}
\ee
which yields eq.~(\ref{beta}).  

\input{DC-aps.bbl}
\end{document}

%% file: DC-aps.bbl
\providecommand{\noopsort}[1]{}\providecommand{\singleletter}[1]{#1}%